\documentclass{amsart}
\usepackage[mathscr]{eucal}
\usepackage{amssymb, amsmath,array, amscd}
\usepackage[OT2,T1]{fontenc}
\usepackage[fontsize=11pt]{scrextend}
\usepackage{tikz-cd} 
\usepackage{graphicx}
\graphicspath{ {./images/} }
\usepackage{adjustbox}
\usepackage{dcolumn}
\usepackage{bm}
\usepackage{graphics}
\usepackage{slashed}
\usepackage{amssymb}
\usepackage{amsmath}
\usepackage{amsthm}
\usepackage{url}
\DeclareFontFamily{T1}{calligra}{}
\DeclareFontShape{T1}{calligra}{m}{n}{<->s*[1.44]callig15}{}
\DeclareMathAlphabet\mathcalligra   {T1}{calligra} {m} {n}
\DeclareMathAlphabet\mathzapf       {T1}{pzc} {mb} {it}
\DeclareMathAlphabet\mathchorus     {T1}{qzc} {m} {n}
\DeclareMathAlphabet\mathrsfso      {U}{rsfso}{m}{n}
\newcommand{\bea}{\begin{eqnarray}}
\newcommand{\ena}{\end{eqnarray}}
\newcommand{\bean}{\begin{eqnarray*}}
\newcommand{\enan}{\end{eqnarray*}}

\makeatletter
\newcommand{\tpitchfork}{%
  \vbox{
    \baselineskip\z@skip
    \lineskip-.52ex
    \lineskiplimit\maxdimen
    \m@th
    \ialign{##\crcr\hidewidth\smash{$-$}\hidewidth\crcr$\pitchfork$\crcr}
  }%
}
\makeatother

\begin{document}

\title[Knotted light]{Towards a unified description of knotted light}


\author[E. Goulart]{E. Goulart}
\address{Departamento de F\'isica e Matem\'atica 
CAP-Universidade Federal de S\~ao Jo\~ao del-Rei, Rod. MG 443, KM 7, 36420-000, Ouro Branco, MG, Brazil}

\email{egoulart@ufsj.edu.br}

\author[J. E. Ottoni]{J. E. Ottoni}
\address{Departamento de F\'isica e Matem\'atica 
CAP-Universidade Federal de S\~ao Jo\~ao del-Rei, Rod. MG 443, KM 7, 36420-000, Ouro Branco, MG, Brazil}

\email{jeottoni@ufsj.edu.br}



\maketitle

\begin{abstract}
Several complementary approaches to investigate knotted solutions of Maxwell's equations in vacuum are now available in literature. However, only partial results towards a unified description of them have been achieved. This is potentially worrying, since new developments of the theory possibly lie at the intersection between the appropriate formalisms. The aim of this paper is to pave the way for a theoretical framework in which this unification becomes feasible, in principle.  
\end{abstract}

\section{Introduction}

The theory of electromagnetic knots continues to spread new insights into the behaviour of electromagnetism in nontrivial regimes (see \cite{Bouwmeester} for a review). Roughly speaking, an electromagnetic knot is a solution of Maxwell's equations in vacuum whose field lines, at any given time, twist in 3-space forming knotted loops which may be pairwise linked. In the particular case of null fields, the corresponding radiative solutions are called \textit{knotted light}. These are characterized by nontrivial helicities and are adapted to an underlying geodesic, shear-free, null congruence, first discussed by I. Robinson \cite{TR,Elbistan,Robinson}. In \cite{Irvine}, Irvine and Bouwmeester have shown how approximate knots of light may be generated using tightly focused circularly polarized laser beams, while recent numerical results were presented in \cite{Numerical}. Also, as discussed in \cite{e Silva}, these solutions can be interpreted in terms of two codimension-2 spacetime foliations, whose leaves intersect orthogonally everywhere in spacetime.

Since A. Ra\~nada's machinery based on a pair of time-evolving Hopf maps  \cite{Ranada1,Ranada2,Ranada3}, several methods have been developed for constructing knotted null fields. Among them the most proeminent are the following: A - Bateman construction, based on the introduction of constrained complex scalar fields satisfying a system of fully-nonlinear first order partial differential equations \cite{Bateman, Besieris}; B - Spinor/twistor method, in which twistor functions correspond to a class of torus knots and the Poynting vector structure forms the spatial part of the Robinson congruence \cite{Kedia,Penrose}; C - Special conformal transformations to generate new solutions from existing ones. This has enabled to get a wide class of solutions
from the basic configurations like constant fields and plane-waves \cite{Sonnenschein}; D - de Sitter method, based on a correspondence of Maxwell solutions on Minkowski and de Sitter spaces, thanks to the conformal equivalence between these spaces and the conformal invariance of four-dimensional gauge theory \cite{Lechtenfeld}.  

It has been pointed out that, since the known methods evolved largely independently, they also suggest different directions in which to expand the study of knotted light \cite{Bouwmeester}. However, to the best of our knowledge, no systematic approach to unify them in a rigorous theoretical framework exist in literature. In order to shed some light on the debate, we think new ways to view interconnections between existing formalisms are very welcome. The aim of this paper is to discuss a framework where new interconnections emerge. In particular, using the machinery of quaternions, parallelizations and Maurer-Cartan forms, we show how the Bateman pairs and Ra\~nada complex maps are tied together. As a consequence, we show how to obtain several physical quantities of relevance.

The paper is divided as follows: in section II we summarize null electromagnetic fields and set concepts to be used throughout. In section III we review the appropriate quaternionic machinery, introduce Lie frammings, Maurer-Cartan forms, adjoint maps and derive important relations. Section IV presents some results in a complex representation, whereas section V uses these results to establish a connection between Bateman pairs and Ra\~nada complex maps. We conclude with some of possible developments that can be explored.

\section{Null electromagnetic fields}

Let $(M,\ g)$ denote Minkowski spacetime with signature convention $(1,3)$. The task of finding null solutions of the source-free Maxwell's equations in an open set $\mathcal{U}\subseteq M$ may be stated as follows. Find a nonzero complex 2-form $R\in\Omega^{2}(\mathcal{U},\mathbb{C})$ such that
\begin{equation}\label{conditions}
\star R=iR,\quad\quad\quad R\wedge R=0,\quad\quad\quad dR=0,
\end{equation}
with $\star$ denoting the Hodge star operator with respect to g $(\star\star=-1)$. The self-duality condition guarantees that $R=F-i\star F$ with $F\in\Omega^{2}(\mathcal{U},\mathbb{R})$: we identify $F$ and $\star F$ with the Faraday tensor and its dual. The second condition says that $R$ is totally decomposable i.e., it splits as an exterior product of complex 1-forms: this automatically defines a null field since this top-degree identity is proportional to the invariants $F_{ab}F^{ab}$ and $F_{ab}\star F^{ab}$. Finally, the closedness of $R$ implies that $F$ satisfies Maxwell's equations in vaccum. We shall call $R$ a \textit{Riemann-Silberstein 2-form} since the interior product $X\lrcorner R$, for a normalized timelike vector field $X$, is essentially $\textbf{E}+i\textbf{B}$.

Let us briefly recall some known properties of Riemann-Silberstein 2-forms (see \cite{Holland} for a detailed discussion). Due to purely algebraic considerations, they split as an exterior product
\begin{equation}
R=k^{\flat}\wedge m^{\flat}
\end{equation}
where $k$ and $m$ are mutually orthogonal null vectors (with $k$ real) and $k^{\flat}$, $m^{\flat}$ the canonical contractions with the metric via musical isomorphisms. Formally, the principal null field $k$ is a section of the null cone bundle $\mathcal{N}$ over $\mathcal{U}$, the closedness of $R$ implying that the flow generated by $k$ constitute a shear-free null geodesic congruence: we say that the electromagnetic field is adapted to $k$. Now, at a given spacetime point $x$, the set $k^{\perp}_{x}=\{v\in T_{x}M|g(k,v)=0\}$ defines a 3-dimensional vector space containing $k$ and, associated to $k^{\perp}_{x}$ there is a two-parameter family of real spacelike planes, the \textit{screen spaces} of $k$. The set of all screens at all spacetime points constitute a bundle $k^{\perp}/k$, called the \textit{umbral bundle}. It is clear that if $k(x)$ and $m(x)$ are sections of $\mathcal{N}$ and $k^{\perp}/k$, respectively, the self-duality of $R$ is then related to a rotation by $90^{0}$ of the screen onto itself i.e., $\star{R}=k^{\flat}\wedge im^{\flat}$.

An additional property of Riemann-Silberstein 2-forms is that they induce two codimension-2 foliations of Minkowski spacetime ($\mathrsfso{F}_{1}$ and $\mathrsfso{F}_{2}$). The corresponding involutive distributions are given by
\begin{equation}
\Delta_{1}=\mbox{ker}\ \mbox{Re}(R)\quad\quad\quad \Delta_{2}=\mbox{ker}\ \mbox{Im}(R)
\end{equation}
and the integral manifolds are 2-dimensional embedded surfaces ruled by null geodesics. The family of all such surfaces constitute a particular instance of the $(2,2)$ dual foliations defined in \cite{e Silva}. Following this reference, we shall call the leaves of $\mathrsfso{F}_{1}$ and $\mathrsfso{F}_{2}$, magnetic and electric, respectively. 


The case of knotted light we discuss here, the Hopf-Ra\~nada solution, is a beautiful realization of a $(2,2)$ dual foliation. For this solution, the electromagnetic field is adapted to a twisting congruence
of light rays (the Robinson congruence) and the magnetic/electric
leaves are pairwise linked once. Furthermore, if $u$ is tangent to a magnetic
leaf and $v$ is tangent to an electric leaf, then $g(u, v) = 0$. We shall see that
the anatomy of this solution is intimately related to the internal structure
of quaternions, which permits a direct route from the formalism of Bateman
pairs to Rañada’s complex mappings. Furthermore, our approach reinforces
aspects of knotted light which are similar in spirit to some aspects of mathematical
gauge theory (specially on the construction of instantons).

\section{Quaternionic machinery}


\subsection{Invariant vector fields and Lie framings}

\quad To begin with, we identify the four-dimensional Euclidean space $(\mathbb{R}^{4}, \delta)$ with the associative algebra of real quaternions $\mathbb{H}$ by giving Hamilton's multiplication table for the canonical basis $\{\textbf{e}_{0},...,\textbf{e}_{3}\}$
\begin{equation}
\textbf{e}_{0}\textbf{e}_{\mu}=\textbf{e}_{\mu}\textbf{e}_{0}=\textbf{e}_{\mu},\quad\quad\quad\textbf{e}_{i}\textbf{e}_{j}=-\delta_{ij}\textbf{e}_{0}+\epsilon_{ij}^{\phantom a\phantom a k}\textbf{e}_{k}.
\end{equation}
A typical quaternion and its conjugate then take the form\footnote{Throughout, lower-case Greeks run from 0 to 3 and lower-case Latins run from 1 to 3.}
\begin{equation}
q=q^{\mu}\textbf{e}_{\mu}=q^{0}\textbf{e}_{0}+q^{i}\textbf{e}_{i}\quad\quad\quad \bar{q}=q^{\mu}\bar{\textbf{e}}_{\mu}=q^{0}\textbf{e}_{0}-q^{i}\textbf{e}_{i}
\end{equation}
and there follow $\overline{pq}=\bar{q}\bar{p}$. As usual, the real part of a quaternion is identified with $\mathbb{R}\textbf{e}_{0}$ whereas the imaginary part is identified with the vector quantity $\mathbb{R}\textbf{e}_{1}+\mathbb{R}\textbf{e}_{2}+\mathbb{R}\textbf{e}_{3}$. The norm of a quaternion $||\ ||:\mathbb{H}\rightarrow\mathbb{R}^{\geq 0}$ is defined by $||q||=\sqrt{q\bar{q}}$ whereas the inverse reads as $q^{-1}=\bar{q}/||q||^{2}$. In what follows, $\mathbb{H}^{*}$ denotes the Lie group of non-zero real quaternions while the unit three-sphere $\mathbb{S}^{3}$ is interpreted as the compact subgroup
\begin{equation}\nonumber
\mbox{Sp(1)}=\{u\in\mathbb{H}^{*}|\ ||u||=1\}.
\end{equation}
Since a generic quaternion may be written as $q=||q||u$ with $u\in\mathbb{S}^{3}$, one sees that $\mathbb{H}^{*}=\mathbb{R}_{+}\times\mbox{Sp(1)}$. Hence, $\mathbb{H}^{*}$ is foliated by the level sets of $||q||$, which are nothing but real three-spheres centered at the origin.
 
Now, as $\mathbb{H}^{*}$ is parallelizable, the left and right translations $l_{q}, r_{q}:\mathbb{H}^{*}\rightarrow \mathbb{H}^{*}$ permit the construction of two globally defined \textit{Lie framings}: the left invariant framing $\varphi_{-}=\{L_{0},...,L_{3}\}$ and the right invariant framing $\varphi_{+}=\{R_{0},...,R_{3}\}$. Specifically, we take
\begin{equation}
L_{\alpha}(q)=q\textbf{e}_{\alpha}\quad\quad\quad R_{\alpha}(q)=\textbf{e}_{\alpha}q 
\end{equation}
and a direct calculation yields the vector fields
\begin{eqnarray*}
&&L_{1}=(-q^{1},q^{0},q^{3},-q^{2})\quad L_{2}=(-q^{2},-q^{3},q^{0},q^{1})\quad L_{3}=(-q^{3},q^{2},-q^{1},q^{0})\\
&&R_{1}=(-q^{1},q^{0},-q^{3},q^{2})\quad R_{2}=(-q^{2},q^{3},q^{0},-q^{1})\quad R_{3}=(-q^{3},-q^{2},q^{1},q^{0}) 
\end{eqnarray*}
with $L_{0}=R_{0}=(q^{0},q^{1},q^{2},q^{3})$. It is easily checked that these fields satisfy the algebraic relations
\begin{equation}
L_{\alpha}\cdot L_{\beta}=||q||^{2}\delta_{\alpha\beta}\quad\quad\quad R_{\alpha}\cdot R_{\beta}=||q||^{2}\delta_{\alpha\beta}  
\end{equation}
where $\cdot$ is the Euclidean scalar product in $(\mathbb{R}^{4},\ \delta)$ and that the Lie brackets read as
\begin{equation}
[L_{\alpha},L_{\beta}]=c_{\alpha\beta}^{\phantom a\phantom a\gamma}L_{\gamma},\quad\quad\quad [R_{\alpha},R_{\beta}]=-c_{\alpha\beta}^{\phantom a\phantom a\gamma}R_{\gamma},\quad\quad\quad [R_{\alpha},L_{\beta}]=0,
\end{equation}
with structure constants given by $c_{ij}^{\phantom a\phantom a k}=2\epsilon_{ijk}$ and all others equal to zero. Consequently, one concludes that the radial field commutes with all other fields and that the framings $\varphi_{\pm}$ automatically induce two complementary parallelizations $\varphi_{\pm}|_{\mbox{Sp(1)}}$, when restricted to the unit 3-sphere. The latter are independent cross-sections of the bundle of orthonormal frames $O(\mathbb{S}^{3})$.


\subsection{Maurer-Cartan forms, co-frames and rank-2 foliations}

In order to deepen on the structure of the invariant vector fields it is illuminating to work within the dual picture. Since left and right invariant fields lead basically to the same theory, we shall concentrate our analysis on the left invariant case. To do so, we define the left invariant Maurer-Cartan one-form
\begin{equation}\label{AV}
\lambda\equiv q^{-1}dq=\lambda^{\mu}\otimes\textbf{e}_{\mu}
\end{equation}
which satisfy, by construction, the relations $\langle \lambda,L_{\alpha}\rangle=\textbf{e}_{\alpha}$. Morally we can think of the Maurer-Cartan form as a compact way to write the standard left-parallelization of the tangent bundle of $\mathbb{H}^{*}$. Direct calculation, using Eqs. (\ref{AV}), give
\begin{equation}\label{generic}
\lambda=d(ln||q||)\otimes\textbf{e}_{0}+||q||^{-2}(q^{0}dq^{i}-q^{i}dq^{0}-\epsilon^{i}_{\phantom a jk}q^{j}dq^{k})\otimes\textbf{e}_{i},
\end{equation}
and writting the quaternion as $q=||q||u$, there follow
\begin{equation}\label{unitary}
\lambda=d(ln||q||)\otimes\textbf{e}_{0}+(u^{0}du^{i}-u^{i}du^{0}-\epsilon^{i}_{\phantom a jk}u^{j}du^{k})\otimes\textbf{e}_{i}
\end{equation}
which implies in 
\begin{equation}
\mbox{Re}(\lambda)=d(ln||q||),\quad\quad\quad \mbox{Im}(\lambda)=\bar{u}du.
\end{equation}
From now on, we shall stick to Eq. (\ref{unitary}) instead of Eq. (\ref{generic}) since it leads to simpler expressions. 

Using the identity
\begin{equation}
d\bar{u}=-\bar{u}du\bar{u}
\end{equation}
one shows that $\lambda$ satisfies the Maurer-Cartan structure equations \cite{Nakahara,Isham} 
\begin{equation}
d\lambda=d\ \mbox{Im}(\lambda)=-\mbox{Im}(\lambda)\wedge \mbox{Im}(\lambda)=-\frac{1}{2}[\mbox{Im}(\lambda)\wedge \mbox{Im}(\lambda)].
\end{equation}
We then read off the relations 
\begin{equation}\label{MC}
d\lambda^{0}=0,\quad\quad\quad\quad d\lambda^{i}=-(\epsilon^{i}_{\phantom a jk}\lambda^{j}\wedge\lambda^{k}),
\end{equation}
implying that each non-vanishing component of the differential is a totally decomposable (simple) 2-form. Notice that although the Maurer-Cartan form itself is $\mathbb{H}$-valued their exterior derivatives take values in $\mbox{Im}(\mathbb{H})\cong\mathfrak{sp}(1)_{L}$. Explicit calculation, using Eq. (\ref{unitary}), gives
\begin{eqnarray}
&& d\lambda=(2du^{0}\wedge du^{i}-\epsilon^{i}_{\phantom a jk}du^{j}\wedge du^{k})\otimes\textbf{e}_{i}.
\end{eqnarray}

An interesting property of $d\lambda$ is that it induces three left-invariant codimension-2 foliations on $\mathbb{H}^{*}$. Indeed, as a consequence of Eq. (\ref{MC}), we have:
\begin{equation}
d\lambda=-2(\lambda^{2}\wedge\lambda^{3}\otimes\textbf{e}_{1}+\lambda^{3}\wedge\lambda^{1}\otimes\textbf{e}_{2}+\lambda^{1}\wedge\lambda^{2}\otimes\textbf{e}_{3}).    
\end{equation}
Defining the rank-2 distributions
\begin{eqnarray}\label{dist1}
D^{i}&\equiv&\mbox{ker} (d\lambda^{i})=\mbox{ker}\lambda^{j}\cap \mbox{ker}\lambda^{k}=\mbox{span}(L_{0}, L_{i})
\end{eqnarray}
with  $i\neq j \neq k$, one may easily check that the latter are integrable in the sense of Frobenius \cite{Lee, Wald}. This means that each 2-plane field define integral manifolds. The associated foliations $\mathcal{F}^{i}$ are transversal to $\mbox{Sp}(1)$ and, therefore, one may interpret the integral curves of the invariant vector field $L_{i}$ (restricted to $\mbox{Sp}(1)$) as intersections of the corresponding foliation with the latter.

It is well known that these curves are linked great circles on $\mathbb{S}^{3}$, called Clifford parallels (or Hopf circles). Collectively, the Clifford parallels comprise three orthogonal Hopf fibrations of $\mathbb{S}^{3}$. The fibers carry Clifford's name because W. K. Clifford discovered them before H. Hopf was born. However, while Clifford understood the fibration quite well, he did not go on to consider the quotient map. Also, one can check that right multiplication yields a framing in which the integral curves of any vector field in the framing are Hopf circles with $+1$ pairwise linking, along which the other two vector fields in the framing spin once in a positive sense (negative sense for the left-invariant fields). Consequently, $R_{i}$ and $L_{i}$ lie in adjacent homotopy classes \cite{Wilson}.

\begin{figure}[h]
    \centering
    \includegraphics[scale=0.6]{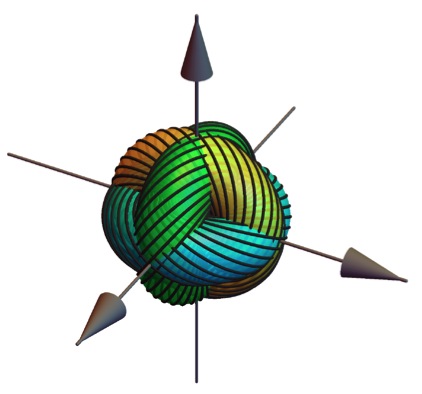}
    \caption{Stereographic projection of $\mathbb{S}^{3}$ permits a visualization of the fibrations. The latter are mutually orthogonal everywhere in $\mathbb{R}^{3}$.}
    \label{fig:ortho}
\end{figure}


\subsection{The adjoint map and leaf description}

In order to describe the 2-dimensional leaves of $\mathcal{F}^{i}$ explicitly,  we start by recalling that $\mathbb{H}^{*}$ acts on itself via the adjoint map
\begin{equation}\label{ad}
ad:\mathbb{H}^{*}\times \mathbb{H}^{*}\rightarrow \mathbb{H}^{*},\quad\quad\quad ad(q,q')=qq'q^{-1}=uq'\bar{u}.
\end{equation}
Clearly, this action leaves $\mbox{Im}(\mathbb{H})$ as an invariant subspace and if $q'$ belongs to the equator of $\mbox{Sp}(1)$ i.e.,  the intersection of the 3-plane spanned by the units $\textbf{e}_{i}$ with the unit three-sphere, $ad(q,q')$ automatically induces a map from $\mathbb{H}^{*}$ to $\mathbb{S}^{2}\subseteq\mbox{Sp}(1)$. Specifically, we consider the quaternionic maps
\begin{equation}\label{adjoint1}
\textbf{l}_{j}(q)=u\textbf{e}_{j}\bar{u}=\textbf{e}_{i}Q^{i}_{\phantom a j}(q)
\end{equation}
where 
\begin{equation}
Q^{i}_{\phantom a j}=\left\{(u_{0}^{2}-u_{k}u^{k})\delta^{i}_{\phantom a j}+2u^{i}u_{j}-2u_{0}\epsilon^{i}_{\phantom a jk}u^{k}\right\}
\end{equation}
with $u_{\mu}=\delta_{\mu\nu}u^{\nu}$. In matrix notation, one obtains
\begin{equation}\label{matrixQ}
Q =\begin{pmatrix}
u_{0}^{2}+u_{1}^{2}-u_{2}^{2}-u_{3}^{2} & 2(u_{1}u_{2}-u_{0}u_{3})&2(u_{1}u_{3}+u_{0}u_{2}) \\
2(u_{1}u_{2}+u_{0}u_{3}) & u_{0}^{2}-u_{1}^{2}+u_{2}^{2}-u_{3}^{2} & 2(u_{2}u_{3}-u_{0}u_{1})\\
2(u_{1}u_{3}-u_{0}u_{2}) & 2(u_{2}u_{3}+u_{0}u_{1}) & u_{0}^{2}-u_{1}^{2}-u_{2}^{2}+u_{3}^{2}
\end{pmatrix}.
\end{equation}

By direct calculation one shows that $Q^{T}Q=\textbf{1}$ and $\mbox{det}(Q)=1$, hence $Q\in\mbox{SO}(3)$. It is easy to show that the integral manifolds associated to the distributions $D^{i}$ are given by the inverse images of the $\textbf{l}_{i}$. Indeed, the differential maps 
\begin{equation}
d\textbf{l}_{i}:T_{q}\mathbb{H}^{*}\rightarrow T_{ad(q,\textbf{e}_{i})}\mathbb{S}^{2}
\end{equation}
are given as follows
\begin{eqnarray}\label{diff1}
&&d\textbf{l}_{i}=du\textbf{e}_{i}\bar{u}+u\textbf{e}_{i}d\bar{u}=u[\lambda,\textbf{e}_{i}]\bar{u}=2\lambda^{j}\otimes\epsilon_{ji}^{\phantom a\phantom a k}\textbf{l}_{k}
\end{eqnarray}
and one obtains
\begin{eqnarray}
&&\langle d\textbf{l}_{i},L_{m}\rangle=2\epsilon_{mi}^{\phantom a\phantom a k}\textbf{l}_{k}\quad\quad\rightarrow\quad\quad L_{0},L_{i}\in\mbox{ker}(d\textbf{l}_{i})
\end{eqnarray}
which proves the assertion.

A remarkable property of the Maurer-Cartan forms $\lambda^{i}$ is that their differentials descend to $\mathbb{S}^{2}$. The claim is that there is a normalized 2-form $\Omega_{2}$ on $\mathbb{S}^{2}$ whose pullbacks along the maps described above define the exterior derivatives $d\lambda^{i}$. Start with the normalized volume form on $\mathbb{S}^{2}$ given by
\begin{equation}
\Omega_{2}=\frac{1}{8\pi}n\cdot (dn\wedge dn)=\frac{1}{8\pi}\epsilon_{ijk}n^{i}dn^{j}\wedge dn^{k}
\end{equation}
with $n=n^{i}\textbf{e}_{i}\in\mbox{Im}(\mathbb{H})$ and $||n||=1$. The pullback of $\Omega_{2}$ along the maps $\textbf{l}_{i}$ read as (no summation)
\begin{eqnarray}
\textbf{l}_{i}^{*}\Omega_{2}=\frac{1}{8\pi} \textbf{l}_{i}\cdot (d\textbf{l}_{i}\wedge d\textbf{l}_{i})
\end{eqnarray}
Using Eq. (\ref{diff1}), we get
\begin{eqnarray*}
\textbf{l}_{i}^{*}\Omega_{2}&=&\frac{1}{2\pi}\lambda^{p}\wedge\lambda^{q}\otimes\epsilon_{ip}^{\phantom a\phantom a k}\epsilon_{iq}^{\phantom a\phantom a l}\textbf{l}_{i}\cdot(\textbf{l}_{k}\textbf{l}_{l})\\
&=&\frac{1}{2\pi}\lambda^{p}\wedge\lambda^{q}\otimes\epsilon_{ip}^{\phantom a\phantom a k}\epsilon_{iq}^{\phantom a\phantom a l}\textbf{l}_{i}\cdot(u\textbf{e}_{k}\textbf{e}_{l}\bar{u})\\
&=&\frac{1}{2\pi}\lambda^{p}\wedge\lambda^{q}\otimes\epsilon_{ip}^{\phantom a\phantom a k}\epsilon_{iq}^{\phantom a\phantom a l}\epsilon_{kl}^{\phantom a\phantom a m}\delta_{im}\\
&=&\frac{1}{2\pi}\epsilon_{ipq}\lambda^{p}\wedge\lambda^{q}
\end{eqnarray*}
where we have used the fact that $\textbf{l}_{i}\cdot \textbf{l}_{j}=\delta_{ij}$. From these relations we get the alternative representation of the differentials
\begin{equation}\label{altrep}
d\lambda^{i}=-2\pi\textbf{l}_{i}^{*}\Omega_{2}. 
\end{equation}
With this expression, the calculus of the Hopf invariants associated to the left-invariant vector fields $L_{i}$ restricted to $\mathbb{S}^{3}$ are straightforward. Indeed, restricting the three maps $\textbf{l}_{i}$ to $\mathbb{S}^{3}$ it is easily seen that they define three proper submersions and we have, by definition:
\begin{equation}\label{hopfinv}
H(\textbf{l}_{i})=\int_{\mathbb{S}^{3}}\omega^{i}\wedge d\omega^{i}\quad\quad (\mbox{no summation})
\end{equation}
with $d\omega^{i}=\textbf{l}_{i}^{*}\Omega_{2}$. Using (\ref{altrep}) and noticing that $\Omega_{3}=\lambda^{1}\wedge\lambda^{2}\wedge\lambda^{3}$ is the invariant volume form on $\mathbb{S}^{3}$, we get $H(\textbf{l}_{i})=-1$. We shall see in Section V that Eqs. (\ref{altrep}) together with Eqs. (\ref{MC}) form the basic clue for connecting the formalism of Bateman's pairs with Ra\~nada's complex mappings approach. In order to do this we first describe some relevant quantities in complex form.

\section{Complex representation}

Although $\mathbb{H}$ is not properly a complex algebra, nonethless, for our purposes it is convenient to represent $\mathbb{H}$ as a real algebra over $\mathbb{C}^{2}$ (see \cite{Delpenich}). In order to do this we revert to the classical notation $\{1,i,j,k\}$ for the principal units and consider the decomposition $\mathbb{H}=\mathbb{C}\oplus j\mathbb{C}$. A typical quaternion, its conjugate and the Hermitian scalar product then take the form
\begin{equation}\label{dec}
q=z_{1}+jz_{2}\quad\quad\quad \bar{q}=\bar{z}_{1}-jz_{2}\quad\quad\quad ||q||^{2}=z_{1}\bar{z}_{1}+z_{2}\bar{z}_{2}
\end{equation}
in which we have introduced the complex coordinates $z_{1}=q^{0}+iq^{1}$ and $z_{2}=q^{2}-iq^{3}$. Since several previous expressions are simpler when written in terms of unit quaternions, we write $u=v_{1}+jv_{2}$ and (\ref{AV}) becomes
\begin{eqnarray}\label{1}
&&\lambda=d(ln||q||)+(\bar{v}_{1}dv_{1}+\bar{v}_{2}dv_{2})+j(v_{1}dv_{2}-v_{2}dv_{1}),
\end{eqnarray}
whereas its exterior derivative is given by 
\begin{eqnarray}\label{2-forms1}
&&d\lambda=(d\bar{v}_{1}\wedge dv_{1}+d{\bar{v}}_{2}\wedge d{v}_{2})+j2(dv_{1}\wedge dv_{2}).
\end{eqnarray}

Writting the matrix $Q$ in Eq. (\ref{matrixQ}) in terms of the complex components, we obtain the quaternionic maps:
\begin{eqnarray*}
&&\textbf{l}_{1}=(v_{1}\bar{v}_{1}-v_{2}\bar{v}_{2})i+j2i\bar{v}_{1}v_{2}\\
&&\textbf{l}_{2}=(v_{1}v_{2}-\bar{v}_{1}\bar{v}_{2})+j(\bar{v}_{1}^{2}+v_{2}^{2})\\
&&\textbf{l}_{3}=(v_{1}v_{2}+\bar{v}_{1}\bar{v}_{2})i+ji(v_{2}^{2}-\bar{v}_{1}^{2})
\end{eqnarray*}
It is convenient for our purposes, however, to perform a stereographic projection
\begin{equation}
\sigma:\mathbb{S}^{2}\setminus (-1,0,0)\rightarrow\mathbb{CP}^{1}\setminus \infty\cong\mathbb{C}
\end{equation}
onto the second copy of $\mathbb{C}$. In order to obtain the most appealing geometrical results we use
\begin{equation}
n\mapsto \zeta=\frac{n^{2}-in^{3}}{1+n^{1}}
\end{equation}
and there follow the composite maps $\zeta_{i}:\mathbb{H}^{*}\rightarrow\mathbb{C}$ 
\begin{eqnarray}\label{cmappings}
&&\zeta_{1}=i\frac{v_{2}}{{v}_{1}}\quad\quad\quad\ \zeta_{2}=\frac{(\bar{v}_{1}+i{v}_{2})}{({v}_{1}+i\bar{v}_{2})}\quad\quad\quad \zeta_{3}=i\frac{(v_{2}-\bar{v}_{1})}{(\bar{v}_{2}+{v}_{1})}
\end{eqnarray}
where $\zeta_{i}=\sigma\circ \textbf{l}_{i}$. Therefore, in order to describe explicitly one of the foliations defined by Eqs. (\ref{dist1}), simply take the appropriate map and consider its preimages. Clearly, each leaf is a complex line through the origin in $\mathbb{H}^{*}$.
Interestingly, since the normalized volume form on $\mathbb{S}^{2}$ may be written as $\Omega_{2}=\frac{1}{2\pi i}d\bar\zeta\wedge d\zeta/(1+\zeta\bar{\zeta})^{2}$ one must have, as a consequence of Eqs. (\ref{altrep}) and (\ref{cmappings}) the identities (no summation)
\begin{equation}\label{dlambda}
d\lambda^{j}=i\frac{d\bar\zeta_{j}\wedge d\zeta_{j}}{(1+\zeta_{j}\bar{\zeta}_{j})^{2}}
\end{equation}

\section{Connection of Bateman pairs and Ra\~nada complex maps}
In order to reproduce the Hopf-Ra\~nada solution using the co-framings described so far we proceed very much in the same way as gauge theorists. To begin with, we let $P(M,\mbox{Sp(1)})$ denote the principal $\mbox{Sp(1)}$ bundle over Minkowski spacetime $M$ and consider a section $s:M\rightarrow P(M,\mbox{Sp(1)})$, defined by
\begin{equation}
u(t,x)=\alpha(t,x)+j\beta(t,x),
\end{equation}
with $\alpha,\beta\in\mathcal{F}(M, \mathbb{C})$ and $|\alpha|^{2}+|\beta|^{2}=1$. As usual, we require that points at spatial infinity in $M$ are mapped to the same point in $\mbox{Sp(1)}$, for all times. To be specific, we assume $u(t,\infty)=\textbf{e}_{0}$, which effectively compactifies spacetime to the product manifold $\mathbb{R}\times\mathbb{S}^{3}$, and $s$ can be thought of as some sort of `gauge function'. The pullback of the Maurer-Cartan form along $s$ defines a $\mathfrak{sp}(1)_{L}$-valued one-form in spacetime
\begin{equation}
s^{*}\lambda=(\bar{\alpha}d\alpha+\bar{\beta} d\beta)+j(\alpha d\beta-\beta d{\alpha}).
\end{equation}
Notice that the first term in the r.h.s is purely imaginary since $\alpha\bar{\alpha}+\beta\bar{\beta}=1$. The exterior derivative is 
\begin{equation}
d(s^{*}\lambda)=(d\bar{\alpha}\wedge d\alpha+d\bar{\beta}\wedge d\beta)+2j(d \alpha\wedge d\beta).
\end{equation}

Now, suppose that we manage to find a smooth section satisfying the fully nonlinear system of first order partial differential equations
\begin{equation}\label{fully}
\star(d\alpha\wedge d\beta)=i(d\alpha\wedge d\beta)\neq 0.
\end{equation}
Connection with the electromagnetic field comes with the identification
\begin{equation}\label{identification}
R=2d\alpha\wedge d\beta,
\end{equation}
where $(\alpha,\beta)$ plays the role of Bateman pairs and the Faraday tensor is given, as before, by the real part of $R$. Since the complex 2-form $R$ qualifies as a Riemann-Silberstein 2-form it may be written as $R=k^{\flat}\wedge m^{\flat}$ and one obtains, after simple manipulations:
\begin{eqnarray}\label{k}
&& k^{\flat}=-i(\bar{\alpha}d\alpha+\bar{\beta}d\beta),\\\label{m}
&& m^{\flat}=2i(\alpha d\beta-\beta d\alpha).
\end{eqnarray}
It can be checked that these quantities satisfy
\begin{equation}
g(k,k)=0\quad\quad g(m,m)=0\quad\quad g(k,m)=0    
\end{equation}
as expected. Consequently, one sees that $k(x)$ and $m(x)$ are necessarily sections of $\mathcal{N}$ and $k^{\perp}/k$, respectively. Due to Robinson theorem \cite{Robinson}, $k$ is geodesic and shear free.

With the above conventions, there follow also the useful relations
\begin{equation}
dk^{\flat}\wedge k^{\flat}\neq 0,\quad\quad dk^{\flat}=im^{\flat}\wedge\bar{m}^{\flat}/4,\quad\quad dm^{\flat}=2ik^{\flat}\wedge m^{\flat}
\end{equation}
from which one concludes that $k$ is a rotating congruence and $R=d(m^{\flat}/2i)$. Therefore, the formalism provides a set of \textit{internal} relations which reveals a rich mathematical structure, connecting at once several quantities of physical relevance. In order to further explore such connections one uses the diagram

\hspace{-0.55cm} 
\adjustbox{scale=1.5, center}{
\begin{tikzcd}
\mbox{Sp}(1) \arrow{r}{\textbf{l}_{i}} \arrow{rd}{\zeta_{i}}  & \mathbb{S}^{2} \arrow{d}{\sigma} \\
M \arrow{r}{\psi_{i}}\arrow[swap]{u}{s}& \mathbb{C}
\end{tikzcd}
}\vspace{1.0cm}
which clarifies the internal anatomy of the solution. Once the Bateman pair is obtained, one uses Eqs. (\ref{cmappings}) to construct the maps 
\begin{equation}\label{Ramaps}
\psi_{1}=i\frac{\beta}{\alpha}\quad\quad\quad\ \psi_{2}=\frac{(\bar{\alpha}+i{\beta})}{(\alpha+i\bar{\beta})}\quad\quad\quad \psi_{3}=i\frac{(\beta-\bar{\alpha})}{(\bar{\beta}+{\alpha})}.
\end{equation}   
Clearly, $\psi_{2}$ and $\psi_{3}$ play the role of the Ra\~nada's maps described in \cite{Ranada1,Ranada2,Ranada3}. Indeed, using Eqs. (\ref{dlambda}), we have:
\begin{eqnarray*}
F=\mbox{Re}(R)=i\frac{d\bar\psi_{2}\wedge d\psi_{2}}{(1+\psi_{2}\bar{\psi}_{2})^{2}},\quad\quad\star{F}=-\mbox{Im}(R)=i\frac{d\psi_{3}\wedge d\bar{\psi}_{3}}{(1+\psi_{3}\bar{\psi}_{3})^{2}}.  \end{eqnarray*}
As discussed in Section II, the preimages of $\psi_{2}$ describes the magnetic leaves whereas the preimages of $\psi_{3}$ describes the electric leaves. As shown in \cite{e Silva}, these are 2-dimensional surfaces embedded in spacetime and ruled by lines defined by $k$. What about the role played by $\psi_{1}$? This map describes the leaves associated to the kernel of the real 2-form
\begin{equation}
S\equiv i m^{\flat}\wedge \bar{m}^{\flat}/4.  \end{equation}
Now, one may complete the vectors $k,m,\bar{m}$ with a real null vector $l$ to construct a null tetrad satisfying:
\begin{equation}
g(k,l)=1\quad\quad g(l,l)=0\quad\quad g(l,m)=0    
\end{equation}
Since $S$ annihilates the distribution spanned by $k$ and $l$, these leaves are defined by a timelike field of 2-planes. If one chooses an observer in this plane, the distribution will contain also the Poynting vector. Therefore, it is natural to call these leaves by light leaves.

It is highly instructive to check how our construction works in the simplest known case: the Hopf-Ra\~nada solution. We recommend the reader to verify the consistency of Eqs (\ref{dlambda}) and (\ref{Ramaps}) in the case of the well-known cross-section
\begin{eqnarray*}
\alpha=\frac{r^{2}-t^{2}-1+2iz}{r^{2}-(t-i)^{2}},\quad\quad\quad\beta=\frac{2(x-iy)}{r^{2}-(t-i)^{2}}
\end{eqnarray*}
with $r^{2}=x^{2}+y^{2}+z^{2}$. For this solution the Riemann-Silberstein 2-form read as
\begin{eqnarray*}
&&R=8iC^{-3}\{[(x-iy)^{2}-(t-z-i)^{2}]\omega^{1}+i[(x-iy)^{2}+(t-z-i)^{2}]\omega^{2}\\
&&\quad\quad\quad\quad -2[(x-iy)(t-z-i)]\omega^{3}\}
\end{eqnarray*}
with
\begin{eqnarray*}
&& C=r^{2}-(t-i)^{2}\\
&&\omega^{1}=dt\wedge dx+idy\wedge dz\\
&&\omega^{2}=dt\wedge dy+idz\wedge dx\\
&&\omega^{3}=dt\wedge dz+idx\wedge dy,
\end{eqnarray*}
and, using Eqs. (\ref{k}) and (\ref{m}), the Robinson congruence (k) and the screens (m) emerge naturally, if we raise indices with the contravariant metric.

\section{Conclusion}
Knotted and linked fields have been investigated in a variety of physical systems: hadron models, topological MHD, classical/quantum field theories, DNA topology, nematic liquid crystals, fault resistant quantum computing among others. In this paper we propose a framework in which a unified description of knotted light becomes transparent.

Since it is well known how to translate the spinor/twistor formalism into Bateman pairs \cite{Kedia}, we have focused our attention on interconnections between the latter and Ra\~nada's complex maps. Starting with a general description of null fields in terms of Riemman-Silberstein 2-forms, we discuss some bundles of physical interest and review some results concerning foliations presented in \cite{e Silva}. We then move to the notion of a parallelization on $\mathbb{H}^{*}$ and derive several useful relations using the Maurer-Cartan forms and the corresponding structure equations. The latter is an essential ingredient in showing that the differential of the Maurer-Cartan form may be obtained from maps to the 2-sphere. We show the explicit form of these maps and discuss how their preimages define three codimension-2 foliations of $\mathbb{H}^{*}$. Intersections of the corresponding complex curves with the 3-sphere give rise to three orthogonal Hopf fibrations. In order to reproduce the Hopf-Ra\~nada solution in terms of the co-frames thus described we proceed very much in the same way as gauge theorists. We consider a principal $\mbox{Sp}(1)$ bundle over spacetime and consider a section satisfying a system of fully nonlinear PDE's. These equations automatically translates into Bateman pairs. The underlying formalism then give, with simple manipulations: the corresponding Ra\~nada's complex maps plus a map whose preimages describe the 2-diemansional leaves associated to the Poynting vector, the Robinson congruence, the section of the umbral bundle and all related potentials.

We hope that the construction presented here may shed some light in further developments of the theory. In particular, it should be interesting to investigate whether the holomorphic approach $h: \mathbb{H}\rightarrow\mathbb{H}$, which give rise to new Bateman pairs, fits in the scenario discussed in the paper. This could give a definitive proof if there exist topology preserving solutions derived from Seifert fibrations. We shall investigate these questions in a forthcomming paper.

\bibliographystyle{amsalpha}

\end{document}